\documentclass[11pt]{revtex4}

\usepackage[textwidth=14cm,textheight=21.5cm]{geometry}

\usepackage{graphicx}
\usepackage{wrapfig}
\usepackage{comment}
\usepackage{amsmath}
\usepackage{color}
\usepackage{amssymb}

\usepackage{dcolumn}
\usepackage{bm}
\usepackage{verbatim}
\usepackage{comment}
\usepackage{wrapfig}
\usepackage{amsmath}
\usepackage{color}
\usepackage{amssymb}
\usepackage{titlesec}
\usepackage{hyperref}

\newcommand{\degree}{\ensuremath{{{}^\circ}}}

\newcommand{\lsi}{LS~I+61\degree~303}

\begin{document}

\title{Multi-Messenger Astrophysics with IceCube}
\author{Mathieu Ribordy$^{a}$ for the IceCube collaboration$^{b}$}

\affiliation{$^{a}$ High Energy Physics Laboratory, EPFL, CH - 1015 Lausanne}
\affiliation{$^{b}$ http://icecube.wisc.edu}

%\PACSes{\PACSit{00.00}{By the way, which PACS is it, the 00.00? GOK.}
%\PACSit{---.---}{\ldots}}

\begin{abstract}
~\vskip4mm

{\it The cubic kilometer IceCube neutrino telescope now operating
at the South Pole in a near complete configuration observes the
neutrino sky with an unprecedented sensitivity to galactic and
extra-galactic cosmic ray accelerators. Within the multi-messenger
framework, IceCube offers unique capabilities to correlate and contrast
the neutrino sky with the gamma-ray sky and ultra high energy
cosmic rays and complements other indirect and direct dark matter search
programs.  
We review here the status of the experiment and recent selected results. A
discussion of the implications of the observations will be followed by
the prospects of future developments, substantially extending the reach
of the observatory at extremely high energies, in the GZK region and at
low energies enhancing capabilities to probe dark matter and cosmic ray
sources in the southern sky.}

\vskip4mm~

\end{abstract}

\maketitle

\section{Introduction}
The IceCube construction signals the emergence of a new class of gigantic detectors dedicated to the observation of the high energy (HE) neutrino sky. 
IceCube's main goals are the unambiguous identification of the first galactic and extra-galactic cosmic ray (CR) accelerators with the detection of HE neutrinos from point sources and the divulgation of the nature of dark matter (DM) through the observation of a secondary neutrino flux from annihilating DM in our galaxy.
A multi-messenger approach applies particularly well to these research topics:  the knowledge which can be gained by combining and contrasting measurements by means of various astroparticles strongly enhances the physics return for the community.

We introduce the IceCube neutrino telescope in section~\ref{sct:icecube} and present selected results revealing the multi-messenger approach potential in section~\ref{sct:mm} :
Searches for DM annihilation in the galactic halo and for neutrinos from the X-ray binary~\lsi~based on time-dependent neutrino flux predictions in sections~\ref{sct:dm} and~\ref{sct:lsi}; a search for a correlation of HE neutrinos with ultra high energy (UHE) CR and the prospects for UHE neutrino detection in sections~\ref{sct:uheNu} and~\ref{sct:prospects} . IceCube as a probe for CR anisotropies in the Southern sky is presented in section~\ref{sct:aniso}.

\section{The IceCube detector}\label{sct:icecube}
The current layout of the IceCube neutrino observatory consists of an in-ice array of 79 strings deployed at a maximal depth of 2.45 km under the South Pole ice cap instrumenting a km$^3$ of crystal clear ice and a surface air shower array IceTop instrumenting a square km. Since the deployment began in 2004, the detector has been operating in the various configurations, referred to as IC-$xx$, where xx is the number of strings in operation, reflecting the construction status.
The observatory is scheduled for completion in December 2010, after a very successful deployment over the past five years~\cite{icecube}.
The design is meeting performance expectations \cite{Achterberg:2006md}.
The in-ice array will eventually be equipped with 86 strings, a nominal string spacing of 125~m, with a maximal number of 60 digital optical modules (a photomultiplier and electronics for signal digitization, time-stamping and communication with the ground surface)~\cite{:2008ym,Abbasi:2010vc}. 
In the center of the in-ice array, 6 strings more densely instrumented form together with the neighbouring strings a dense inner core, enhancing the IceCube detection capabilities toward lower energies and potentially enabling IceCube to explore the Southern neutrino sky. 

In its current state, IC-79 is taking data at a rate of approximately 2 kHz, dominated by secondary muons originating from cosmic air showers in the atmosphere.  The current integrated exposure is equivalent to about two years with the completed IceCube detector. Moreover, with the now operational deep core, the data are enriched with lower energy neutrinos. 
After the application of filtering and reconstruction procedures to the events~\cite{Ahrens:2003fg}, the major experimental muon background can be largely reduced in order to keep a sample of events dominated by neutrino-induced muon events. Astrophysical searches look for an excess over atmospheric neutrino background expectations.  

The IceTop air shower array is dedicated to composition studies of the CR spectrum around the knee and above \cite{Ahrens:2004nn}. IceTop enables the detection of air showers at energies above $\approx$1 PeV. As a part of IceCube, it also plays an important role of facilitating background rejection and calibration of events triggered by the in-ice array \cite{Ahrens:2004dg}.

On site, prospecting activities for the detection of UHE neutrinos are taking place as well, relying on the alternate radio Cherenkov and acoustic signatures accompanying UHE neutrino interactions, see Sct.~\ref{sct:prospects}.

\section{Multi-messenger Astroparticle Physics with IceCube}\label{sct:mm}
\subsection{Searches for supersymmetric dark matter }\label{sct:dm}
Under certain assumptions, the minimal supersymmetric standard model (MSSM) framework provides a stable weakly interacting DM candidate: the neutralino, a self-annihilating thermal relic of the early universe. Its mass, bounded from below by accelerator constraints and above by theory, lies between 46 GeV~\cite{abda} up to a few TeV~\cite{Griest:2000kj}. Secondary particles, including $\nu$, are emitted at a higher rate from regions of greater DM density where gravitationally trapped neutralinos annihilate pairwise~\cite{carlos}. 

The galactic halo or compact objects such as the Sun seem to be promising regions of such enhanced DM densities and thus for conducting dedicated searches for these signatures. These analyses all search in common for an excess from the directions of these enhanced DM density regions.
In the absence of any excess over the known atmospheric 
neutrino background, upper limits on the neutrino-induced muon flux from DM annihilations are obtained.
The deeper connection to the physics arises with the conversion of the neutrino-induced muon flux limits into cross section upper limits~\cite{Wikstrom:2009kw}: 
self-annihilation cross sections, velocity averaged (in halo analyses), $\sigma_A v$, and spin-dependent scattering cross sections $\sigma_{\mathrm{SD}} $ in search for signatures from the self-annihilating solar neutralinos (assuming equilibrium between capture and annihilation rate in the Sun).
Assuming neutralinos constitute a sizable fraction of the galactic DM density, these analyses have a significant potential to exclude regions from the MSSM parameter space which would otherwise remain unconstrained by direct search experiments~\cite{Halzen:2009vu} and by $\gamma$-ray and CR experiments performing indirect searches similar to IceCube~\cite{Abbasi:2009uz}.
\begin{figure}[here]%{r}{6.5cm}
\centering
\includegraphics*[scale=.31]{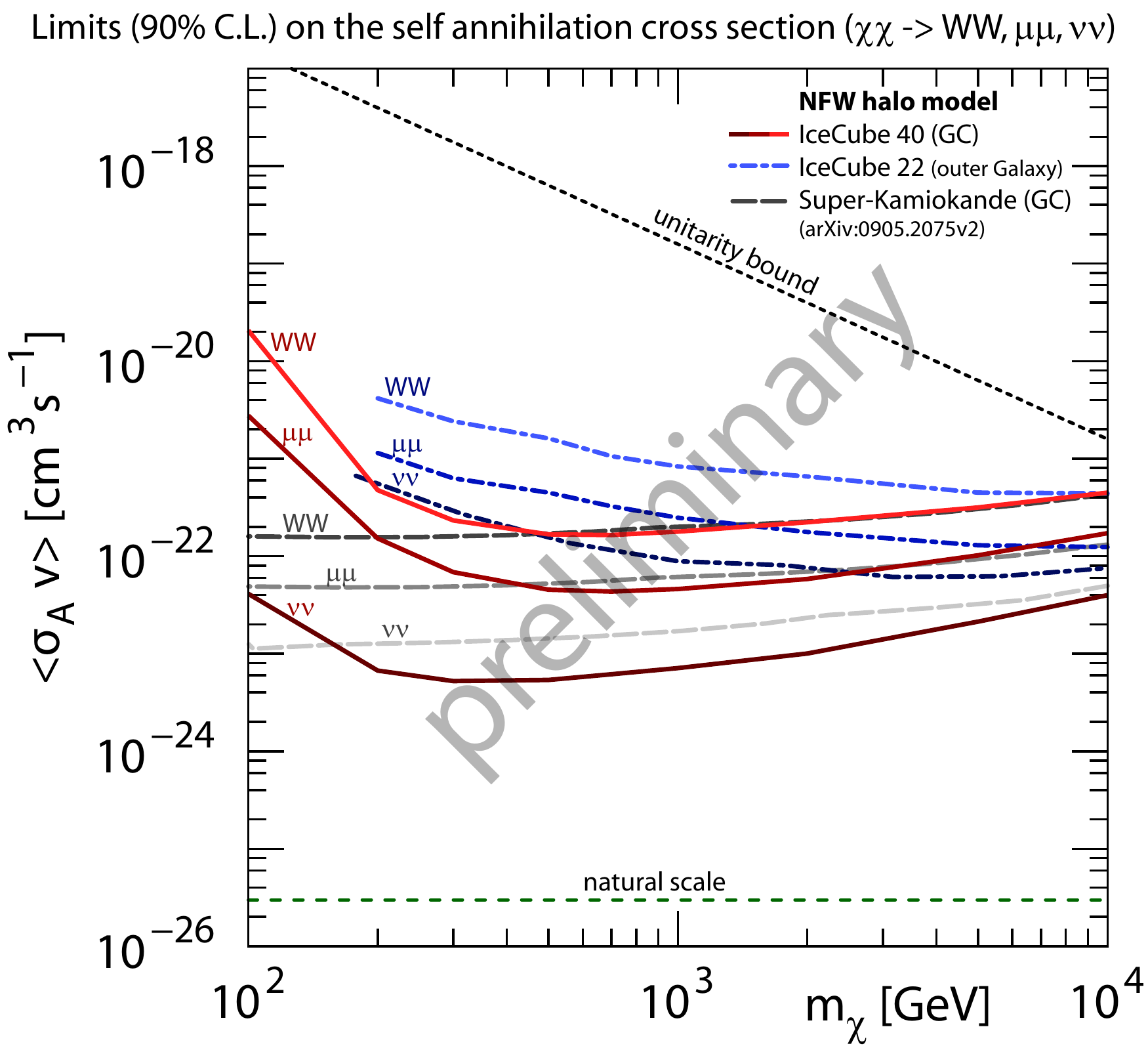}
\includegraphics*[scale=.21]{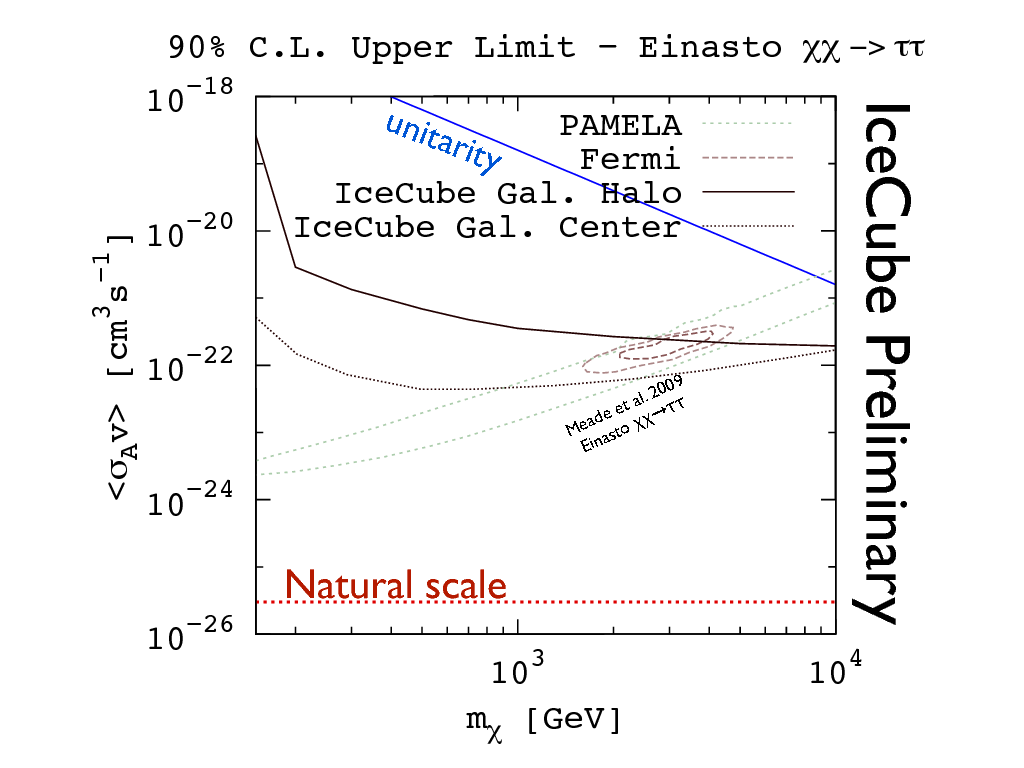}
\caption{Left: limits on $\sigma_A v$ w.r.t. $m_\chi$ for various self-annihilation channels and both halo analyses, including SK results for comparison~\cite{Desai:2004pq,Hisano:2009fb}. Right:  limits of the halo analyses and favored region by PAMELA and Fermi for the $\tau\tau$ self-annihilation channel.}
\label{fig:dm-limits}
\end{figure}

Novel results from searches for a DM signature from the halo are presented here as they seem promising; indeed powerful as well as competitive and complementary with other indirect searches. Two different searches were performed, respectively with IC-22 and IC-40~\cite{carsten}. 
The first search looked for a differential excess from the northern sky selecting two broad regions, one pointing toward the Galactic Center (but not including it, as it is located in the Southern hemisphere) and a corresponding mirror region (defined by a shift of 180\degree in right ascension). This analysis is referred to as the {\it outer Galaxy analysis}.
The second and more ambitious is a search for an excess from the Galactic Center (GC); a difficult undertaking due to the heavy atmospheric muon background contamination as the GC is above the horizon (i.e. in the Southern hemisphere). The search is made possible for the first time with IC-40 thanks to the conjunction of a large detection volume enabling to veto the background efficiently and the fact that the signal, according to specific models profiling the DM density in the Milky Way~\cite{DMprofile}, is strongly increasing in this direction (the neutralino annihilation rate scales with the square of the DM density~\cite{Yuksel:2007ac}). This analysis is however slightly more dependent on the assumed DM profile, uncertain near the GC.

Results of these IceCube's searches are presented in Fig.~\ref{fig:dm-limits}. From a multi-messenger perspective they are remarkable: combining the measurements from the Fermi and PAMELA experiments and in a leptophilic ($\chi\chi\rightarrow\tau\tau$) neutralino self-annihilation scenario~\cite{meade}, the favored region is excluded by IceCube's GC analysis (considering the Einasto profile). This is a compelling illustration of the great contrasting potential of observations, which is achieved by combining observational constraints issued by means of the various messengers.
Note however, that the direct superposition on a single plot of results, which rely on the validity of the assumed DM profile, must be interpreted with care: the neutrino flux exclusion limit comes from the integrated self-annihilating DM rate along the line of sight, while the anomalous measurements of the positron fraction by PAMELA and of the electronic component by Fermi signs local annihilation, as electrons and positrons cool off quickly. A more conservative interpretation of the anomalous measurements calls for local CR accelerators~\cite{Hooper:2008kg,Hu:2009zzb,Ackermann:2010ip,yuksel}.

\subsection{Search for neutrinos from the X-ray binary \lsi}\label{sct:lsi}
While the resolution of the centennial mystery of CR origin constitutes one of IceCube main's goals as it would deepen our understanding of the astrophysical sources which accelerate them, hadronic accelerators have not yet been revealed. Tight constraints have instead been issued with an average upper flux limit from a random excess in the northern sky which lies at $E^2 {\mathrm{d}}\Phi/{\mathrm{d}}E \approx 0.5\cdot 10^{-11}$ TeV cm$^{-2}$ s$^{-1}$ (IC-40) for Fermi-accelerated CR in steady point source of neutrinos~\cite{Abbasi:2010ki}. 
This limit can be substantially improved however at the expense of an increased model dependency of the search, for instance by looking at an excess from predefined sources or source class (source stacking) in a catalog~\cite{Abbasi:2010ki,Achterberg:2006ik}.
This limit can be further overcome by accounting for time-dependence of the neutrino emission in the test statistics with an assumed neutrino flux enhancement related to the variability of the multi-wavelength spectrum~\cite{Braun:2009wp}:

1. If the source is not periodic and exhibits random MWL variability, e.g. blazar, analyses assuming a positive correlation between specific photon bands and the neutrino flux were conducted~\cite{Abbasi:2010ki}. These searches are strongly model-dependent.

2. If the source is periodic, e.g. the gamma-ray loud X-ray binary \lsi ~ considered below, we have considered two approaches: either assuming an equally periodic enhancement of the neutrino emission, but unknown normalization and phase as described in~\cite{mike-theseProc} or, alternatively,  applying a new analysis methodology, based on specific models relating multi-wavelength spectra  to the expected spectral neutrino "light curves"~\cite{levent}. While this second approach suffers a reduced statistical penalty (the price to pay for the fit of the parameters) and may thus boost the IceCube's discovery potential, it makes the search more strongly model-dependent.

We explore here this new analysis methodology and apply it to the fascinating gamma-ray loud X-ray binary ~\lsi, a periodic neutrino source candidate using the hadronic model described in~\cite{Chernyakova:2009hp,Neronov:2008bw} among others~\cite{Christiansen:2006ac, Romero:2005fr, orellanaRomero,romero,Torres:2006ub} and recently subject to intense multi-wavelength observation campaigns~\cite{MAGIC, VERITAS,Tanaka:2007zzc, Torres:2010wm,Smith:2008kq}.
Together with LS 5039 and PSR B1259-63, located in the Southern hemisphere and therefore out of the field of view of IceCube, the TeV emitter \lsi~ exhibits a strong component in the GeV band, variability over two timescales, the orbital period with 26.5 days and the superorbital period with 4.6 yrs, also a notable source of outburst~\cite{gregory}. The nature of the compact object is not yet known~\cite{romero} and the related inclination of the system orbital plane w.r.t. the line of sight not well constrained~\cite{Casares:2005wn}. The massive companion is a Be star with a surrounding decretion disk periodically disrupted over the superorbital timescale, suggesting a possible modulation of the neutrino emission. Near the apastron, the source exhibits enhanced VHE, X-ray and radio activity~\cite{Zabalza:2010fw,paredes97,MAGIC,VERITAS} anti-correlated with the GeV emission~\cite{fermiGeVlsi}.

\begin{figure}[here]
\centering
\includegraphics*[trim = 0mm 0mm 0mm 160mm,clip,scale=.3]{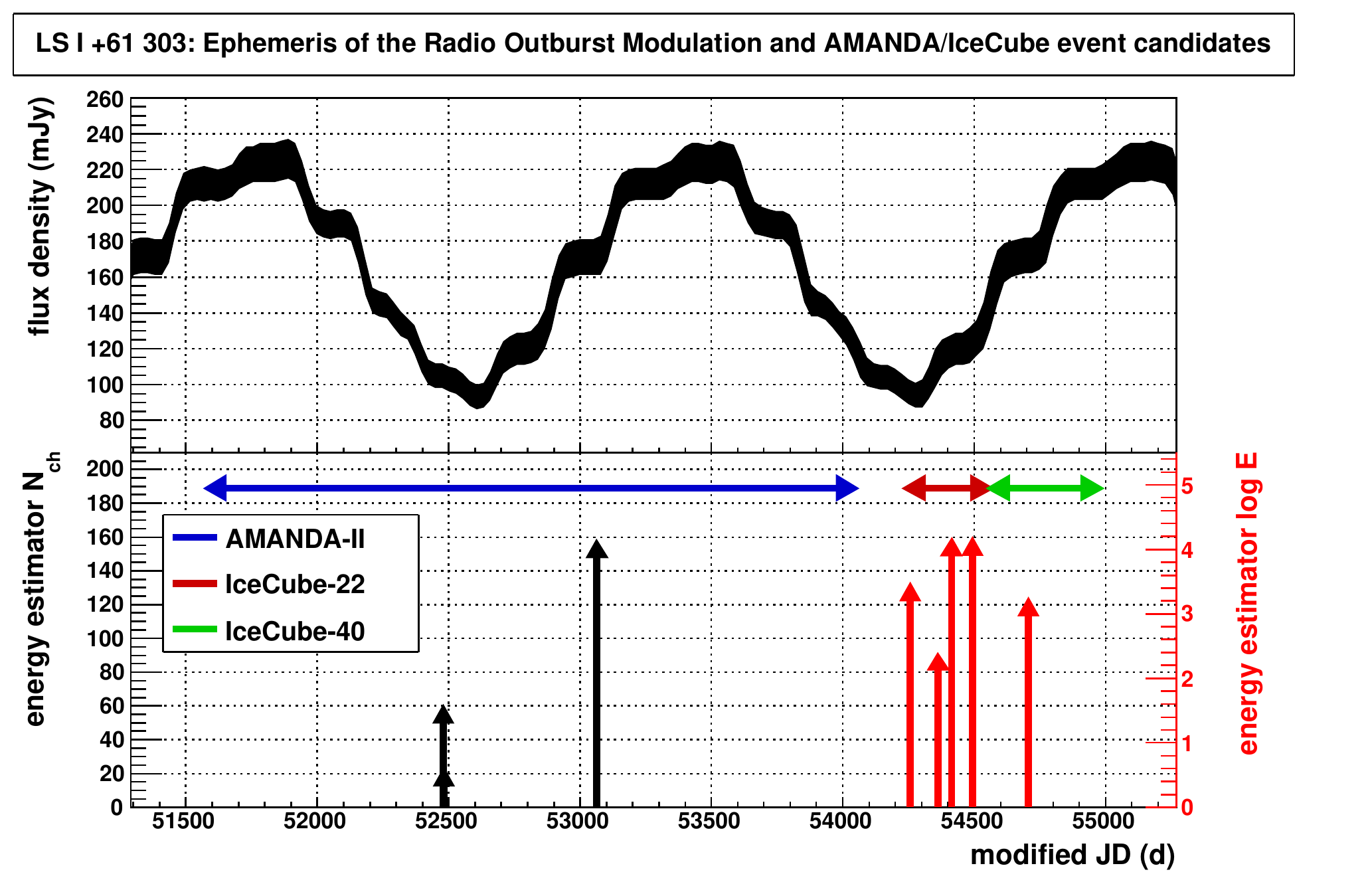}
\includegraphics*[trim = 0mm 0mm 0mm 160mm,clip,scale=.3]{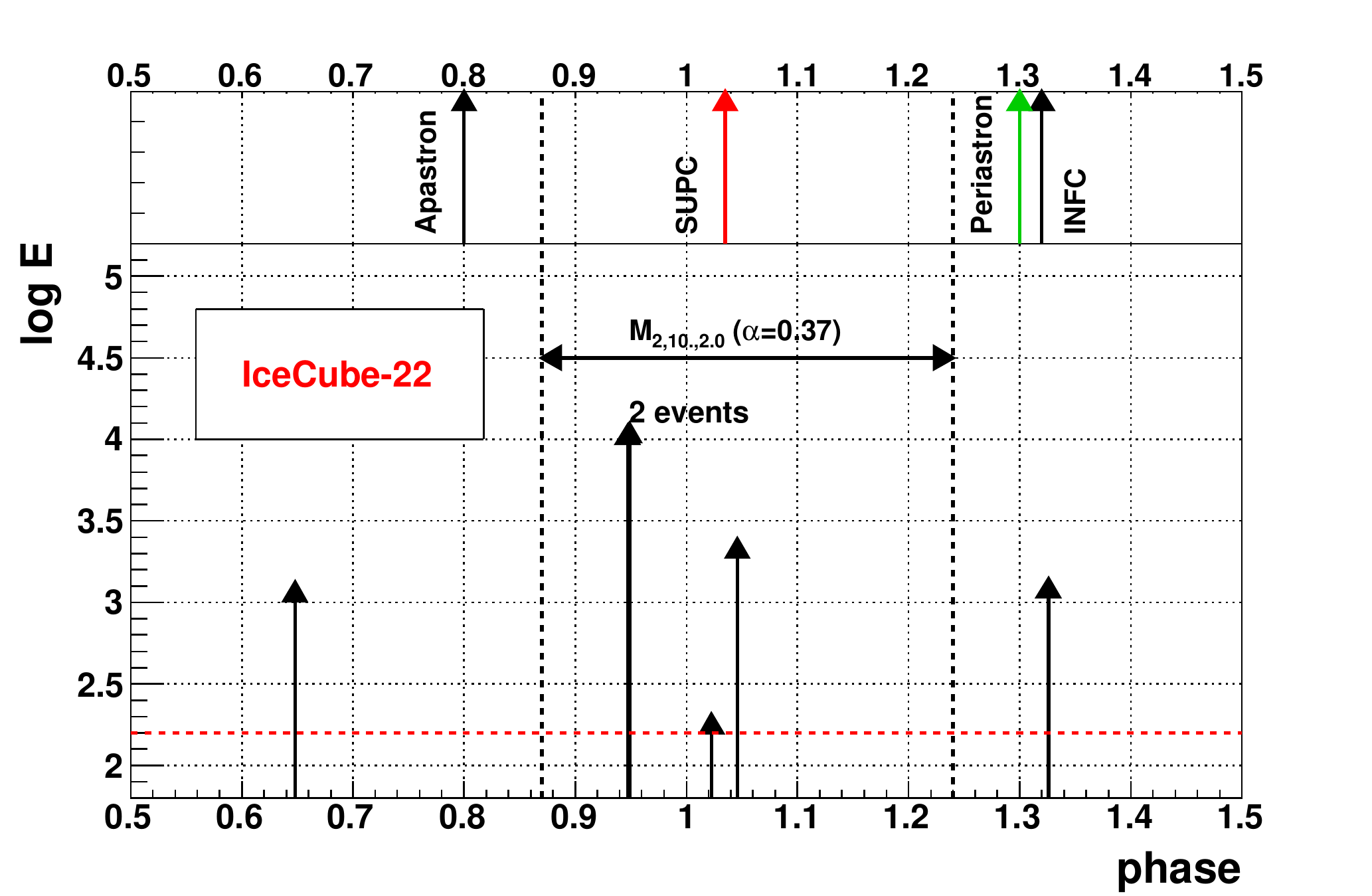}
\caption{Left: selected neutrino candidate events with AMANDA, IC-22 and IC-40 and the superobital radio modulation within the emission window of a model (see text).
Right: phase and energy estimate of the selected events in IC-22.}
\label{fig:FIG_IC22_final_events}
\end{figure}

Following an analysis optimizing the discovery potential, candidate neutrino events were selected from IC-22 and IC-40. 
An event subsample from the publicly released AMANDA events~\cite{amandaPublicEvent} was selected using a similar analysis. 
Fig.~\ref{fig:FIG_IC22_final_events}, left, shows the final event sample recorded by AMANDA, IC-22 and IC-40 in the last 10 years together with the radio modulation. On the right, the phase of the IC-22 selected events is presented. 
The  a-posteriori p-value of about 1\% is obtained from the HE AMANDA events.
The p-value of the IC-22 analysis is approximately 2\%.
The p-value extracted from the blind IC-40 search is not significant. Refined models for neutrino flux predictions including a mechanism to explain the superorbital phase (such as a Be star disk growth followed by a disruption) and the radio and X-ray variabilities on this time scale could be of interest for a long term data analyses of the neutrinos from the direction of ~\lsi.
In conclusion, this approach set more stringent constraints on neutrino emission models and may facilitate the discovery and characterization of a neutrino signal from galactic and extra-galactic point sources. Upon discovery, not only a first incontrovertible demonstration of hadronic acceleration in the environment of specific astrophysical systems would be provided, but the underlying source acceleration mechanisms and morphology would be constrained. 

\subsection{Ultra high energy cosmic rays and neutrinos}\label{sct:uheNu}
The Pierre Auger Observatory (PAO) and HiRes experiments have  in the past decade reported 35 events above 57 EeV, pointing back to the northern hemisphere~\cite{Abraham:2007si}. At these high energies, UHE CRs likely travel only weakly deflected over long extra-galactic distances, thus approximately point back to the cosmic accelerator from which they originate. Therefore, associated neutrinos produced either locally or along the propagation are closely aligned, motivating a search for a directional correlation between neutrinos and UHE CRs.
Moreoever  the strong correlation of the UHE CR arrival directions with nearby active galactic nuclei which was earlier reported by PAO makes this analysis exciting. With the release of an updated UHE CR catalog~\cite{:2010zzj}, the association now has a reduced significance, but there are still indications for the possible existence of such a correlation.

An analysis was conducted with a subsample of the IC-22 data, enriched with HE neutrinos (with a mean energy 100 TeV for an $E^{-2}$ simulated spectrum): the neutrino candidate events were selected on the basis of their angular distance $\Psi<3\,\degree$ to one of the 35 reported UHE CR events. 60 events were found, while 43.7 events were expected (``off-source'' estimation). The probability of such an excess is 0.98\% in the background-only hypothesis. A similar analysis was repeated with the IC-40 data, including the events from the updated PAO catalog, bringing the number of UHE CR events to 82: the previous excess is partially washed out (298 events observed, 274 expected).

\subsection{Alternative GZK neutrino detection technique}\label{sct:prospects}
The IceCube potential for the observation of GZK~\cite{gzk} neutrinos strongly depends on the GZK flux normalization, which depends on source evolution, injection spectrum and CR composition~\cite{Engel:2001hd}.
The characterization of the GZK neutrino flux would enable the partial recovery of the degraded information carried by UHE CR, which would permit the delineation of cosmological source evolution scenarios from source injection spectrum characteristics. This in turn would help determine the nature of the most powerful CR accelerators in the universe.

Currently, the situation concerning the GZK neutrino flux normalization is uncertain~\cite{Ahlers:2010fw,Anchordoqui:2007fi}: while the observed correlation of UHE CR sources with the AGN distribution by AUGER~\cite{Abraham:2007si} hints at a light composition (and in this case current neutrino flux limits lie close to the upper flux predictions~\cite{Abbasi:2010ak}), dedicated AUGER composition studies favor a composition becoming heavier at UHE~\cite{Kampert:2009zz}.
It is therefore necessary to guarantee an observation of GZK neutrinos (and perhaps to characterize its flux) to build detectors with substantially larger effective volumes than IceCube. However, these detectors should not rely on the optical signature, as this detection technique has practical limitations linked to a rather short attenuation length for Cherenkov light propagation in the ice~\cite{sp-ice}. Alternative detection techniques are called for~\cite{aska:a,aska:c,aska:d,learned:1979,bevan:2007,Dedenko:1997ur}:
A rather promising technique is based on the observation of the radio Cherenkov signature of the interaction shower ~\cite{Zas:1991jv}. Over the past decade the ice has been well characterized with the RICE detector and the attenuation length of the radio signal was demonstrated to be much longer than for the optical signal~\cite{rice}, correspondingly enabling a sparser instrumentation of the detection volume. The Askaryan Radio Array (ARA)~\cite{ara} project is now approved for a 3-year preparatory phase, aiming at demonstrating the feasibility of the full scale detector from technological and environmental aspects. ARA would eventually be instrumenting 80 km$^2$  and be providing a 3-year sensitivity sufficient for the detection of GZK neutrinos induced by photo-disintegration in pessimistic predictions of pure iron composition. 
An alternative technique relies on detection of the acoustic signature of the GZK neutrinos. Its applicability is studied by the South Pole Acoustic Test Setup (SPATS)~\cite{spats}. It was recently found that the attenuation length is not as long as anticipated~\cite{attlen,price}, challenging the design of the acoustic detection devices~\cite{acousticRandD}. The experimental effort will continue with the deployment of additional devices this winter at the South Pole aiming to measure the absolute level of environmental noise, which may, if too high, impede the applicability of the technique in this location. The technique, yet to mature from technological and environmental characterization standpoint, could eventually be used in conjunction with the radio detection technique:  the detection of a hybrid event, i.e. detected with two or more distinct signatures (optical, radio or acoustic), would unambiguously demonstrate the interaction of a UHE neutrino.

\subsection{Cosmic ray anisotropies}\label{sct:aniso}
\begin{figure}
\centering
\includegraphics*[scale=0.8]{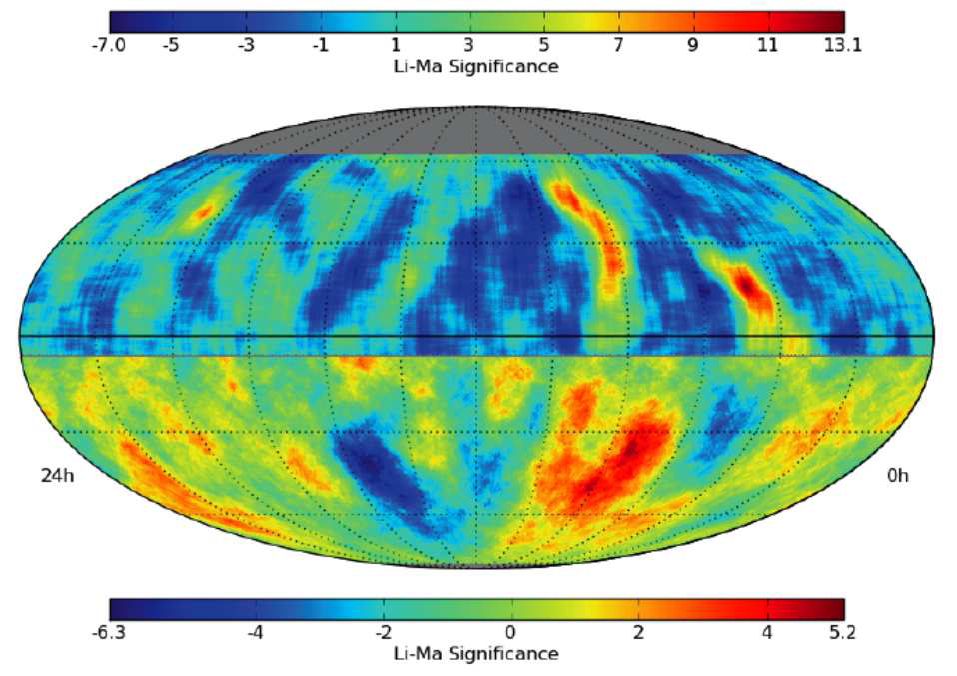}
\caption{Preliminary sky map significance combining data from IC-40 and MILAGRO.} 
\label{fig:aniso}
\end{figure}

Diverting the multi-messenger approach into multi-messenger detection capabilities of the IceCube's in-ice component, we have searched and found first evidence for southern sky anisotropies in the arrival direction of CRs.
Analyses of the IC-22 and IC-40 data are described in detail in~\cite{Abbasi:2010mf, Desiati:2010cz}. The sample of IC-22 down-going muon data reported at this conference shows an anisotropy. The maximal peak to peak amplitude of the projected skymap in right ascension is about 0.15\%. Moreover, when considering two event subsamples with increasing median energies, from 12.6 to 126 TeV, the relative large scale anisotropy reduces significantly.
The combined significance skymap of TeV MILAGRO and multi-TeV IC-40 events is shown Fig.~\ref{fig:aniso} suggest that these medium scale anisotropies could be part of a larger scale anisotropy.

The origin of these anisotropies, reported by several surface and underground experiments located in the northern hemisphere~\cite{crExp, milagro} in the past two decades, is not yet well understood: anisotropies are expected due to the Compton-Getting effect~\cite{compton-getting}, but have not yet been confirmed. On the contrary, the anisotropy seems to be out of phase with this effect. 
The tail-in excess combined with the energy dependence of the observed anisotropy may suggest a local origin, explained by heliospheric effects~\cite{Lazarian:2010sq}, or other local features of the magnetic field over larger scales, which would enable the particles to stream from nearby cosmic accelerators. In this case, it would as well establish a connection with the electronic component anomalies discussed in section~\ref{sct:dm}. 
It is clear from the above that the better characterization of the anisotropies, at various angular scales, could be quite revealing of the involved underlying physics.

\section{Conclusion}
We have presented recent results at the interface of the three pillars of astroparticle physics: particle physics, astrophysics and cosmology, with a clear focus on the multi-messenger approach, which may help delineate the nature of the cosmic accelerators or unveil the nature of the DM. 

IceCube has recently realized its potential for the study of transient events: the IC-40 neutrino flux limits on Gamma-Ray Bursts (see e.g.~\cite{Desiati:2010cz, Abbasi:2009ig}) is now excluding the Waxman-Bahcall upper bound~\cite{wbGrb} and in the future, fully exploiting the deep core in the multi-messenger context, IceCube will certainly be revealed as a powerful tool for issuing strong constraints on the nature of the DM and for the observation of MeV neutrinos from galactic supernova~\cite{Ahrens:2001tz,Halzen:1995ex}.

\acknowledgments
The author, supported by the Swiss National Research Foundation, grant No PP002â 114800, warmly thanks the organizer for a great conference.

\end{document}